\documentclass[preprint,prb]{revtex4}
\usepackage{epsfig}
\usepackage{latexsym}
\usepackage{dcolumn}
\usepackage{color}
\usepackage{subfigure}
\usepackage{bm}
\usepackage{graphicx}
\begin{document}
\date{\today}
\title{Array of Josephson junctions with a non-sinusoidal current-phase
relation as a model of the resistive transition of unconventional
superconductors.}

\author{Anna Carbone}\email{anna.carbone@polito.it}
 \affiliation{Physics Department, Politecnico di Torino,\\ Corso Duca degli
Abruzzi 24, I-10129 Torino, Italy}

\author{Marco Gilli}\email{marco.gilli@polito.it}
\affiliation{Electronic Engineering Department, Politecnico di Torino, Corso Duca degli Abruzzi 24, 10129 Torino, Italia}

\author{Piero Mazzetti}\email{piero.mazzetti@polito.it}
 \affiliation{Physics Department, Politecnico di Torino,\\ Corso Duca degli
Abruzzi 24, I-10129 Torino, Italy}

\author{Linda Ponta}\email{linda.ponta@polito.it}
 \affiliation{Physics Department, Politecnico di Torino,\\ Corso Duca degli
Abruzzi 24, I-10129 Torino, Italy}

\begin{abstract}
An array of resistively and capacitively shunted Josephson junctions with nonsinusoidal  current-phase relation is considered for modelling the transition in high-T$_c$ superconductors.  The emergence of higher harmonics, besides the simple sinusoid $I_{c}\sin\phi$, is expected for dominant \emph{d}-wave symmetry of the Cooper pairs,  random distribution of potential drops, dirty grains, or nonstationary conditions. We show that additional cosine and sine terms act respectively  by modulating the global resistance and  by changing the Josephson coupling of the mixed superconductive-normal states.\\
First, the approach is applied to simulate the transition in disordered granular superconductors with the weak-links characterized by nonsinusoidal current-phase relation.  In granular superconductors, the emergence of higher-order harmonics affects the slope of the transition.  Then, arrays of intrinsic Josephson junctions, naturally formed by the CuO$_2$ planes in cuprates, are considered. The critical temperature suppression, observed at values of hole doping close to $p=1/8$, is investigated. Such suppression, related to the sign change and modulation of the Josephson coupling across the array, is quantified in terms of the intensities of the first and second sinusoids of the current-phase relation. Applications are  envisaged for the design and control of quantum devices based on stacks of intrinsic Josephson junctions.
\end{abstract}
\pacs{74.81.Fa, 74.78.Bz, 74.81.-g, 74.62.-c}
\keywords{Josephson junction arrays, unconventional superconductors}
\maketitle

\section{Introduction}
Arrays of Josephson junctions are under intensive investigation for potential implementation as quantum bits and for modelling several  phenomena in superconductive films \cite{Clark,Caspar,Hilgenkamp,Marconi,Orgiani,Fazio,Zazunov,Refael,Fistul,Shukrinov,Beloborodov}. In particular, the resistive transition has been described by means  of resistively and capacitively shunted Josephson junctions  arrays (RSCJ) \cite{Chakravarty1,Chakravarty2,Harris1,Yu,Ponta}.
In the conventional RSCJ model, the Josephson current is the simple sinusoid  $I_\mathrm{S}(\phi) = I_{c}\sin\phi$, where $I_{c}$   is the
critical current and $\phi = \theta_2-\theta_1$ the phase difference of the superconductor order parameters $\Delta_1 \exp (i\theta_1)$ and $\Delta_2 \exp (i\theta_2)$ . Sign and magnitude of $I_c$ are affected by the gap function symmetry and relative orientation of the superconductor electrodes. For conventional phase-coherent pairing with \emph{s}-wave symmetry and $\Delta_1=\Delta_2=\Delta$,  the critical current is given by the Ambegaokar-Baratoff expression $I_c=\pi \Delta^2/2eR_o \tanh[\Delta/2kT]$, with $R_o$ the normal-state resistance \cite{Josephson}.
  For unconventional high-T$_c$ superconductors, the internal structure of Cooper pairs  most likely agrees with predominant \emph{d}-wave  symmetry that might originate deviations in the superconductive and normal branches of the current-voltage characteristics. Moreover, ferromagnetic impurities, grain boundaries and interfaces, vortex cores, impurities and far-from equilibrium conditions cause the onset of higher harmonics \cite{Tsuei,Kosztin,Han,Bussmann,Il'ichev,Il'ichev2,Mazin,Kuroki,Koch,Kwon,Ivanov,Kopnin,Bezuglyi,Brinkman,Argaman,Gulevich,Binder,Schuster,Frolov,Hikino,Yanase,Ooi,Savelev}.
The current-phase relation is given by:

\begin{equation}\label{Is1}
I_S(\phi) \propto \int_{-\infty}^{+\infty}[ 1-2f(E)]\mathrm{Im}[I_E(\phi)] dE \hspace{5pt},
\end{equation}
with $f(E)$ the electron energy distribution and $\mathrm{Im}[I_E(\phi)]$ the spectral current, which depend on material, geometry and nonequilibrium conditions. Eq.~(\ref{Is1}) can be written as an $n$-order Fourier series
\cite{Golubov,Tanaka1,Tanaka2}:
\begin{equation}\label{Is2}
I_\mathrm{S}(\phi)=\sum_{n \geq 1}\left[{\tilde{I}_{{n}} \sin(n\phi)+\tilde{J}_{{n}} \cos(n\phi)}\right] \hspace{5pt}.
\end{equation}
 When the sum is restricted to $n=1$, $\tilde{I}_{n}\sin(n\phi)$ reduces to the familiar sinusoidal Josephson current $I_c \sin \phi$. $\tilde{J}_{{n}}\cos(n\phi)$ is the quasi-particle-pair-interference  current, vanishing when the pair-symmetry is not broken.
    For \emph{s}-wave superconductors and $n=1$,  $\tilde{I}_{n}$ shows a logarithmic divergence at $V = 2\Delta/e$, whereas $\tilde{J}_{n}$ is zero for $V< 2\Delta/e$ with a discontinuity at $V= 2\Delta/e$ both in normal ($0$) and ferromagnetic ($\pi$) junctions
     \cite{Hikino,Hamilton,Harris,Soerensen77,Pedersen78}. It has been established that higher harmonics are important
in cuprates. At the same time, non-monotonic temperature dependence of
Josephson current also appears in d-wave system \cite{Tanaka2,Barash,Il'ichev3,Testa}.
For unconventional superconductors with prevalent \emph{d}-wave pairing, the harmonic $\sin 2\phi$ is critically enhanced by the presence of midgap Andreev resonant state \cite{Hu,Tanaka3} and can even dominate over $\sin\phi$,
as found in \cite{Il'ichev,Il'ichev2,Lindstrom,Schneider,Zikic}. %
Deviations from the sinusoidal shape can be more easily observed at temperatures below $T_c$ because, in general, these
effects are of the second order. Nonetheless, in the vicinity of $T_c$, they have been observed in normal-metal weak-links, as a consequence
 of the depairing either by proximity effect by supercurrent or in long junctions or in nonequilibrium conditions \cite{Golubov}.
Spin-singlet/spin-triplet superconductor \cite{Asano,Tanaka4} and superconductor/ferromagnet  hybrid structures have triggered considerable interest in recent years for their potential spintronics applications as they allow for tuning the critical current via the electron spin.  If the metal between conventional superconductors is magnetic, the symmetry is broken and the current takes the more general form: $I_s = I_0 \sin(\phi+\phi_0)$. The phase-shift $\phi_0$ is proportional to the magnetic moment perpendicular to the potential of the spin-orbit coupling \cite{Buzdin,Konschelle,Grein,Eschrig}.
\par In this work,  a model of the superconductive-resistive transition based on a network of resistively and capacitively shunted nonsinusoidal
Josephson junctions (RSCNJ) is considered. Such a network could be relevant when the overall and concurring effects of above described phenomena should be taken
into account.  The ultimate scope being the consistent description of several experimental evidences,
that cannot be accounted for by the simple sinusoidal coupling. The appropriateness of the resistively-capacitively picture in the presence of the
nonsinusoidal current-phase relies on the occurrence of the macroscopic quantum tunnelling in high-T$_c$ materials with $d$-wave symmetry,
whose experimental evidence has been reported only very recently \cite{Bauch,Jin}.
\par
Arrays of weak-links in polycrystalline superconductors and  intrinsic Josephson junctions in cuprates are considered as prominent examples.
\par
 In polycrystalline superconductors in the vicinity of the transition, nonequilibrium effects make the relevant properties of the weak-links spatially and temporally dependent on the external drive \cite{Ponta,Kopnin,Bezuglyi,Brinkman,Argaman,Gulevich}.
Hence,  when a polycrystalline superconductor undergoes the transition, the onset of higher harmonics may occur according to the local voltage,
geometry and chemistry of the grains. The pair-interference  current $\tilde{J}_{{n}}\cos(n\phi)$ emerges when the pair-symmetry is broken and comes
into play when the junctions are partly dissipative. Therefore, the role of $\tilde{J}_{{n}}\cos(n\phi)$ may become relevant in the mixed
state close to T$_c$, for current $I\sim I_c$ and voltage $0<V<V_c$.
\par
 Intrinsic Josephson junctions are naturally formed in cuprates and correspond to pairs of CuO$_2$ planes, separated by insulating layers. Such arrays have become attractive for quantum computation  \cite{Bauch,Jin}. Higher harmonics alter the profile of the tilted washboard potential and, thus, the sequence of tunnelling and dissipation processes determining the quantum device operation.
Furthermore, arrays of intrinsic Josephson junctions biased in the resistive
 state where the Josephson current oscillates are being deployed as terahertz emitters \cite{Ozyuzer,Bulaevskii,Tachiki}.  Upon decreasing the bias from the fully resistive state, the emission power increases  as the Josephson frequency resonates with the cavity. With further voltage decreasing, some junctions may fall
 into the superconducting state, thus increasing voltage on the other junctions and, ultimately, switching off the radiation.  Fine tuning  and control of such intertwined oscillating-dissipative processes
  is crucial for the correct operation of the emitter.
\par
The critical temperature anomalies, observed when the $p$-doping of the CuO$_2$ planes is varied, are quantified.
Such anomalies have been related to the emergence of a striped high-T$_c$ phase, with spatially modulated superconducting order,
depending on the doping level $p$. An effective higher-order Josephson coupling varying as a cosine function of twice the difference of
the superconducting phases on adjacent planes has been demonstrated. Several concomitant evidences of antiphase ordering in cuprates,
besides the strong suppression  of T$_c$, have been reported \cite{Liang,Yuli,Homma,Naqib,Berg}. The T$_c$ suppression will be modeled
by using the nonsinusoidal Josephson junctions with the ratio of the second to first harmonics depending on the doping level $p$.
We remark that an array of junctions with simple sinusoidal current-phase, while  correctly describes homogeneous low-T$_c$  superconductors
characterized by uniform positive Josephson coupling, seems quite inadequate for the complex phenomenology of strongly correlated high-T$_c$ cuprates.
%
\section{Nonsinusoidal RSCJ Model}
\begin{figure}
\centering
\includegraphics[width=6cm, angle=0]{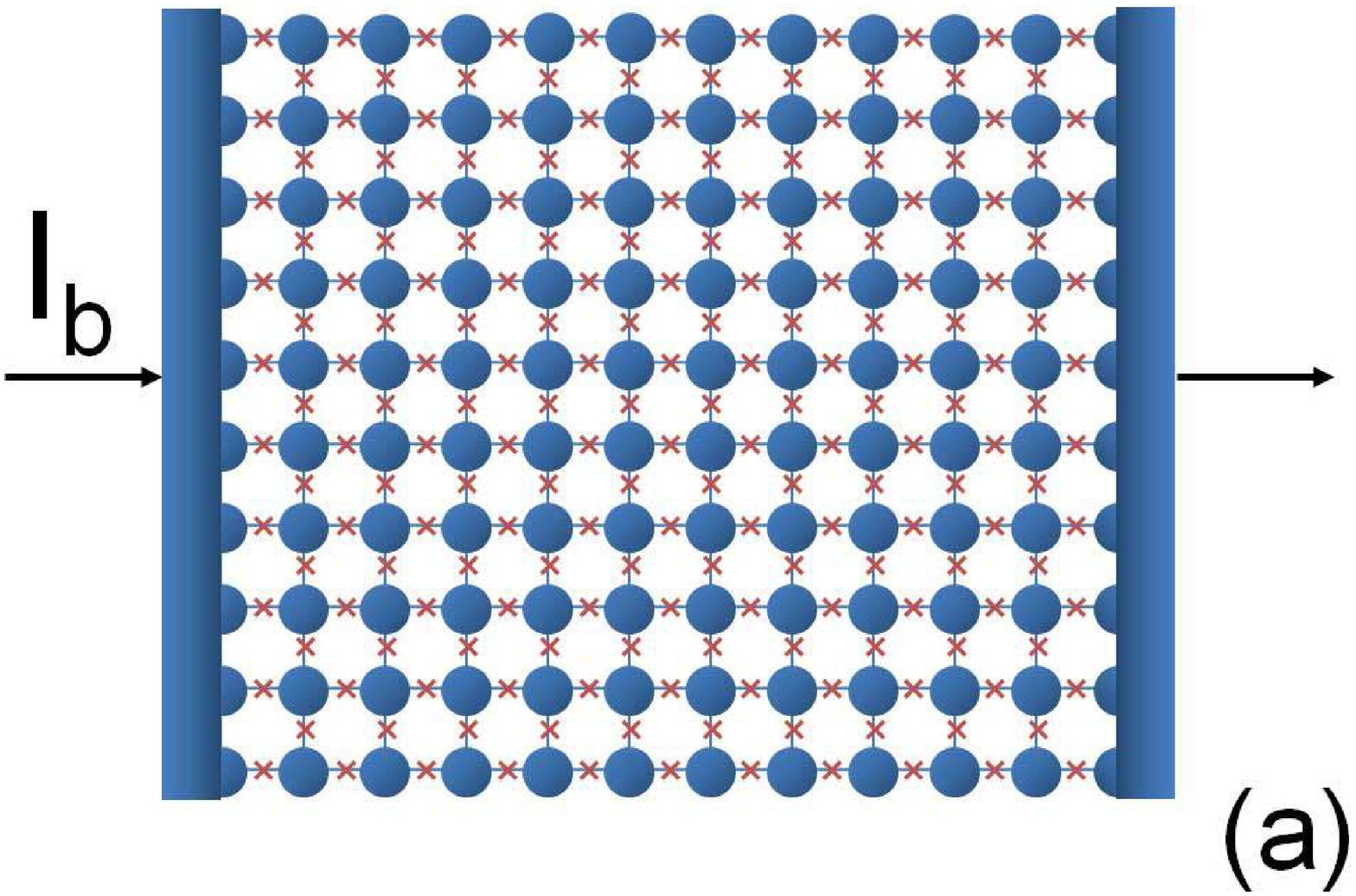}
\includegraphics[width=5cm, angle=0]{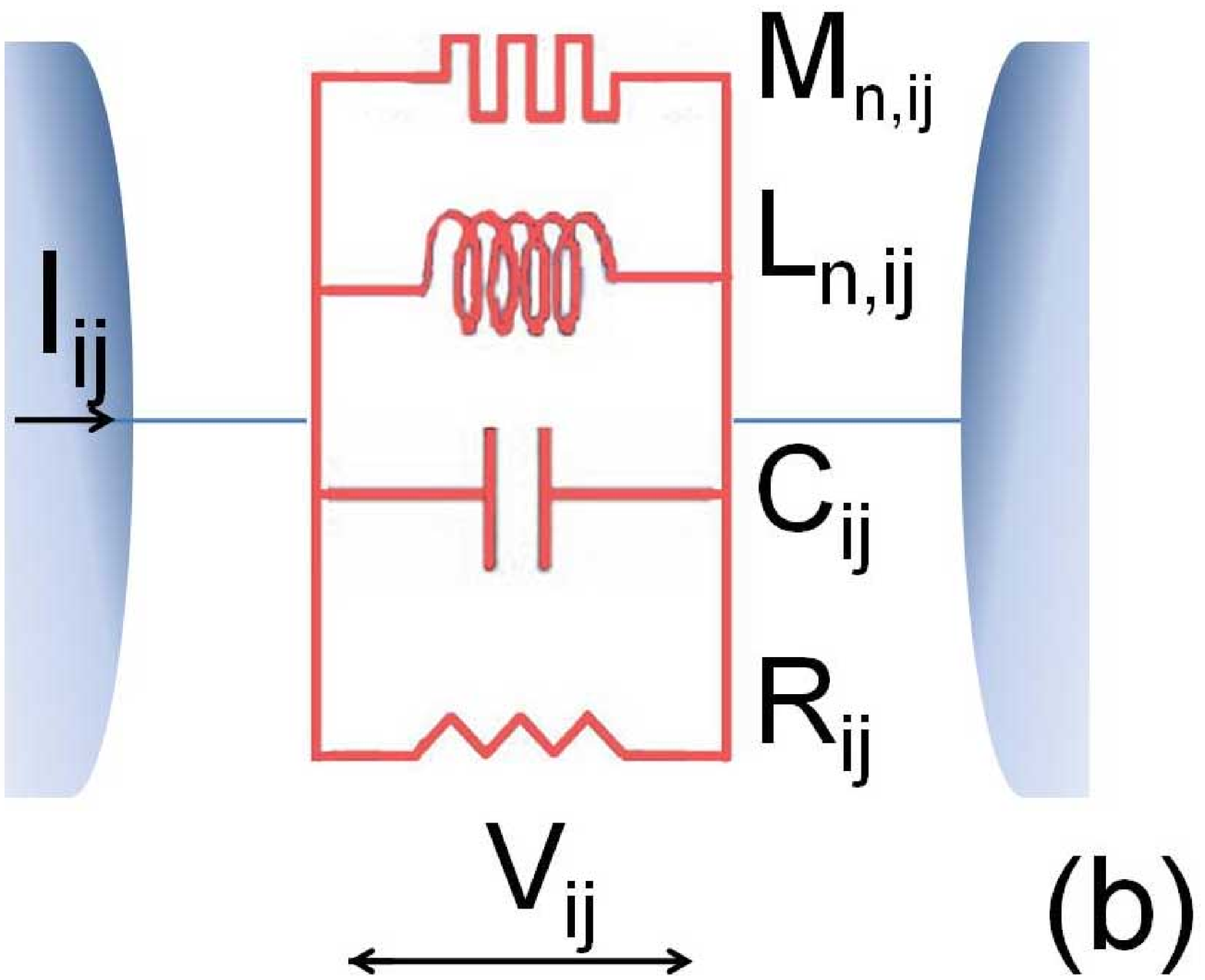}
\caption{\label{Figure1} (a) Two-dimensional Josephson junction array representing a granular superconductor.  Circles represent superconducting grains. Crosses represent weak-links between grains. The bias current $I_b$ is injected to the left electrode and collected from the right electrode. (b) Equivalent circuit of the weak-link between the grains $i$ and $j$.  The linear resistor $R_{ij}$,  the linear capacitor $C_{ij}$, the nonlinear inductor $L_{n,ij}$ and memristor $M_{n,ij}$  are connected in parallel. The current $I_{ij}$ flows from  grain $i$ to grain $j$. $V_{ij}$ is the voltage drop across the weak-link.}
\end{figure}

A two-dimensional array of Josephson junctions is sketched in Fig.~\ref{Figure1}(a). The bias current $I_b$ is injected to the left electrode and collected from the right electrode. Circles represent superconducting grains connected by weak-links (crosses). According to the RSCJ model, the current  $I_{ij}$ flowing through each junction is:
\begin{equation}\label{I}
I_{ij}= C_{ij}\frac{dV_{ij}}{dt}+\frac{V_{ij}}{R}+ I_{\mathrm{S},ij}(\phi_{ij}) + \delta I_{L,ij} \hspace{5pt}.
\end{equation}
where $C_{ij}$ and $R_{ij}$ are the shunt capacitance and  resistance between grains $i$ and $j$,   $I_{\mathrm{S},ij}(\phi_{ij})$ is the Josephson current,  $\delta I_{L,ij}$ is the Langevin fluctuation source. The voltage drop across the junction is given by:
 \begin{equation}
 \label{V}
 V_{ij}= V_i-V_j= \frac{\hbar}{2e}\frac{d \phi_{ij}}{dt} \hspace{5pt},
 \end{equation}
\noindent
with $\phi_{ij}$ the phase difference of the order parameters in the  grains $i$ and $j$.   In the usual RSCJ model, $I_{\mathrm{S},ij}(\phi_{ij})$ is a simple sinusoid, whereas in the present work the nonsinusoidal form given by Eq.~(\ref{Is2}) is considered. Therefore, the current $I_{ij}$ flowing through each junction connecting the grains $i$ and $j$ writes as:
\begin{equation}\label{II1}
    I_{ij}= C_{ij}\frac{dV_{ij}}{dt}+\frac{V_{ij}}{R_{ij}}+\sum_{n \geq 1}[{\tilde{I}_{n,ij} \sin(n\phi_{ij})+\tilde{J}_{n,ij} \cos(n\phi_{ij})}] + \delta I_{L,ij}\hspace{5pt}.
\end{equation}
$I_{ij}$ is given by the sum of the following contributions: the charging current through the shunt capacitance $C_{ij}$, the Ohmic current through the shunt resistance $R_{ij}$, the $n$ Josephson current sources  $\tilde{I}_{n,ij}\sin(n\phi_{ij})$ and $\tilde{J}_{n,ij}\cos(n\phi_{ij})$ and the Langevin current.
\par
 It is worth noting that for $n=1$, $\tilde{I}_{1,ij}\sin \phi_{ij}$ is the familiar sinusoidal Josephson current $I_{c,ij} \sin \phi_{ij}$, whereas $\tilde{J}_{1,ij}\cos(n\phi_{ij})$ with  $\tilde{J}_{{1,ij}}\propto V_{ij}/R_{ij}$ corresponds to the voltage  $V_{ij}$ times a phase-dependent conductance term $ 1/R_{ij} \cos(\phi_{ij})$  \cite{Josephson,Harris,Hamilton,Soerensen77}. Therefore, Eq.~(\ref{II1}) can be rewritten as:

\begin{equation}\label{If1}
    I_{ij}= C_{ij}\frac{\hbar}{2e}\frac{d\phi_{ij}}{dt}+ \frac{1}{R_{ij}}\frac{\hbar}{2e}\frac{d \phi_{ij}}{dt}\cos \phi_{ij}+ I_{c,ij} \sin \phi_{ij} + \delta I_{L,ij}\hspace{5pt},
\end{equation}
\noindent
where Eq.~(\ref{V}) has been used. The second term on the right hand side of Eq.~(\ref{If1}) is commonly called the \emph{interference current}.
\par
The equivalent circuit of a junction obeying Eq.~(\ref{II1}) is shown in Fig.~\ref{Figure1}(b). It corresponds to the parallel of a linear capacitor $C_{ij}$, a linear resistor $R_{ij}$, a parallel of $n$ inductors $L_{n,ij}$ (related to  the $\tilde{I}_{n,ij}\sin(n\phi_{ij})$ terms) and a parallel of $n$ memristors $M_{n,ij}$ related to the $\tilde{J}_{n,ij}\cos(n\phi_{ij})$ terms (we use the notation memristor after \cite{Chua}).

\par
Eq.~(\ref{II1}) can be  written more compactly as:
\begin{equation}\label{IJ}
I_{ij}= C_{ij}\frac{dV_{ij}}{dt}+\frac{V_{ij}}{R}+ \sum_{n \geq 1}I_{c,n,ij}\sin (n\phi_{ij}+\phi_{o,n,ij})+ \delta I_{L,ij} \hspace{5pt},
\end{equation}
with:
\begin{equation}\label{IJ1}
I_{c,n,ij}=\sqrt{\tilde{I}_{n,{ij}}^2+ \tilde{J}_{n,{ij}}^2}\hspace{5pt},
\end{equation}
 and:
 \begin{equation}\label{IJ2}
 \phi_{o,n,ij}=\arctan\left(\frac{\tilde{J}_{n,{ij}}}{\tilde{I}_{n,{ij}}}\right)\hspace{5pt}.
 \end{equation}

\par
 Conventional Josephson junctions are usually classified in terms of the Stewart-McCumber parameter $\beta_c=\tau_{\mathrm{RC}}/\tau_\mathrm{J}$  with  $\tau_{\mathrm{RC}}=RC $ and $\tau_\mathrm{J}=\Phi_o/2\pi I_c R_o$, as overdamped ($\beta_c\ll1$), general ($\beta_c\simeq 1$) and  underdamped ($\beta_c \gg 1$). For the nonsinusoidal junction described by Eq.~(\ref{IJ}), the definition of the Stewart-McCumber parameter can be generalized as follows:
 \begin{equation}\label{SM1}
 \beta_{c}^*=\frac{\tau_{\mathrm{RC}}}{\tau_{\mathrm{J}}^*} \hspace{5pt},
 \end{equation}
 with
 \begin{equation}\label{SM2}
 \tau_{\mathrm{J}}^*=\frac{\Phi_o}{2\pi \sum_n I_{c,n,{ij}}R_o} \hspace{5pt}.
 \end{equation}
 \par
\begin{figure}
\centering
\includegraphics[width=5cm, angle=0]{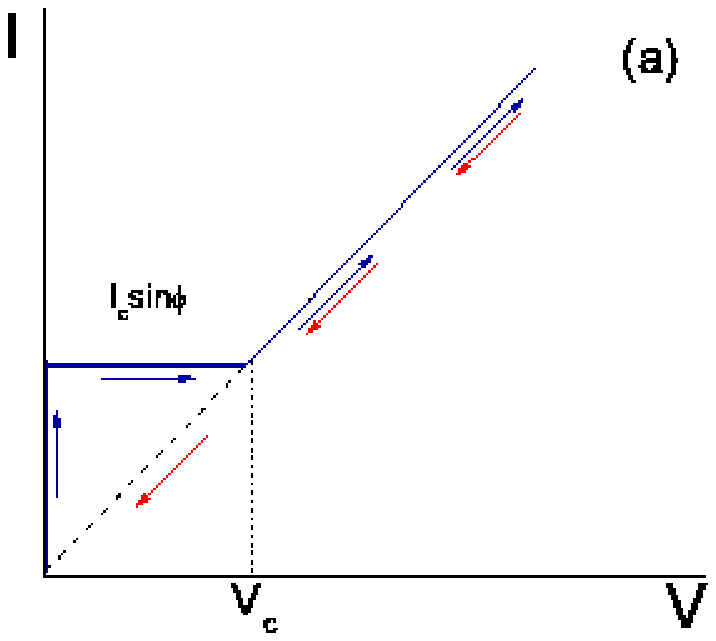}
\includegraphics[width=5cm, angle=0]{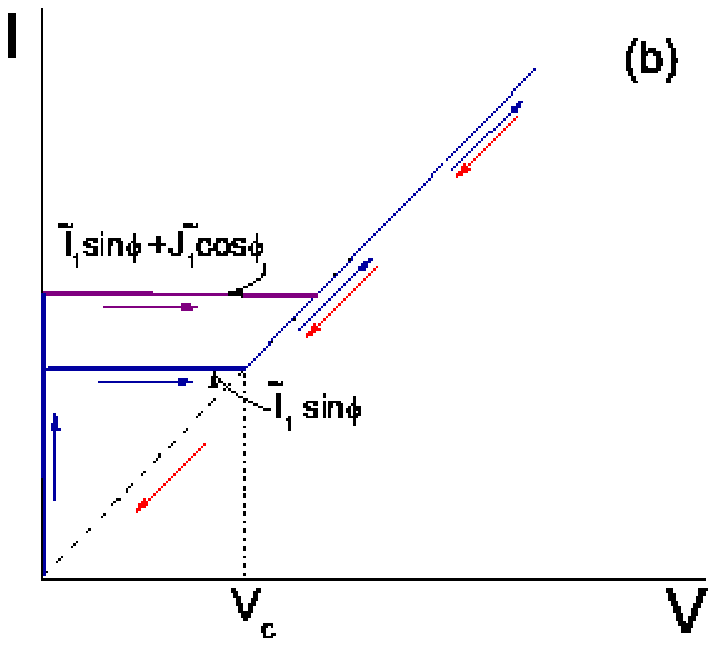}
\includegraphics[width=5cm, angle=0]{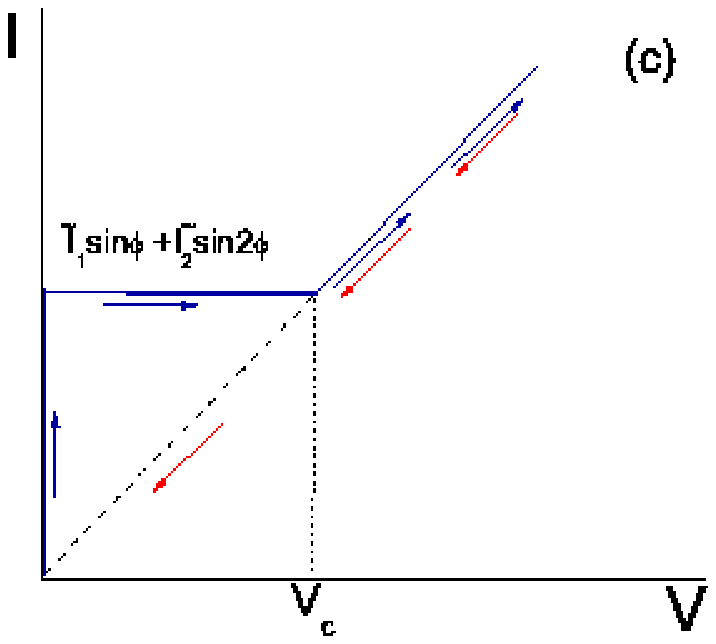}
\caption{\label{Figure2}  Josephson junction characteristics of a weak-link with current-phase relation (a) $I_{\mathrm{S}}(\phi)=I_{c}\sin(\phi)$; (b) $I_{\mathrm{S}}(\phi)=\tilde{I}_{1}\sin(\phi)+\tilde{J}_{1}\cos(\phi)$ with $\tilde{I}_{1}=1\mathrm{mA}$ and  $\tilde{J}_{1}=0.5\mathrm{mA}$; (c) $I_{\mathrm{S}}(\phi)=\tilde{I}_{1}\sin(\phi)+\tilde{I}_{2}\sin(2\phi)$ with $\tilde{I}_{1}=1\mathrm{mA}$ and  $\tilde{I}_{2}=0.5\mathrm{mA}$. The generalized Stewart-McCumber parameter  is $\beta_c^*=45$.}
\end{figure}

 Eq.~(\ref{IJ}) can be numerically solved for an arbitrary number $n$ of harmonics. Nonetheless, we will restrict our discussion to the following two cases relevant for the applications:
  \begin{equation}\label{I1}I_{\mathrm{S},ij}(\phi_{ij})=\tilde{I}_{1,ij}\sin(\phi_{ij})+ \tilde{J}_{1,ij}\cos(\phi_{ij})\hspace{5pt},
  \end{equation}
  and
  \begin{equation}\label{I2}
  I_{\mathrm{S},ij}(\phi_{ij})=\tilde{I}_{1,ij}\sin(\phi_{ij})+ \tilde{I}_{2,ij}\sin(2\phi_{ij})\hspace{5pt}.
  \end{equation}
 The scheme of the current-voltage characteristics of an underdamped ($\beta_c^* \gg 1$) Josephson junction obtained by solving Eq.~(\ref{IJ}) is shown in Fig.~\ref{Figure2}. In particular, Fig.~\ref{Figure2}(a) refers to the simple sinusoid, Fig.~\ref{Figure2}(b) refers to $I_{\mathrm{S},ij}(\phi_{ij})$ given by Eq.~(\ref{I1}) and Fig.~\ref{Figure2}(c) refers to $I_{\mathrm{S},ij}(\phi_{ij})$ given by Eq.~(\ref{I2}). The intermediate states are characterized by voltage drops in the range $0<V_{ij}<V_{c,{ij}}$ and current $I_{ij}=I_{c,n,{ij}}$.  Upon current (voltage) decrease starting from the normal state, the behavior is always resistive, implying that the system reaches the superconductive ground state without exploring the intermediate states. For overdamped junctions ($\beta_c^* \ll 1$), the intermediate states are characterized by voltage drop and current respectively in the range $0<V_{ij}<2V_{c,{ij}}$  and $I_{c,n,ij}<I_{ij}<I_{c,n,ij}[2V_{c,ij}]$. Upon increasing and decreasing the external drive, the current-voltage behavior  is the same, hence no hysteresis is observed. In the general case ($\beta_c^* \approx 1$), the $I-V$ curve is partly hysteretic. Upon increasing the external drive, the intermediate states are characterized by a voltage drop in the range $0<V_{ij}<V_{c,{ij}}$ and current equal to $I_{c,n,ij}$. As the external drive decreases, the backward current lies slightly below the forward current. It is worthy of remarks that  the capacitive effect is reduced with the nonsinusoidal current phase relation in comparison to the simple sinusoidal case.\\
As a final remark, we note that since the simulations are addressed at modeling the zero-frequency (time-asymptotic) response
 of a macroscopic array, the Langevin term does not affect the results and thus in the simulations
the term $\delta I_{L,ij}$ can be set to zero. Nonetheless,  we stress that the term $\delta I_{L,ij}$ has profound conceptual implications related to the microscopic random  dissipation/tunneling events  and the onset of decoherence according
 to the Caldeira-Leggett picture. The term $\delta I_{L,ij}$ plays a major role in the evaluation of current noise power spectra shape and amplitude \cite{Clark}.
\subsection{Resistive transition in granular superconductors}
The resistive transition is modeled by using a disordered network of weak-links with nonsinusoidal current-phase relation. The network is routinely solved by a system of Kirchhoff equations by using  Eq.~(\ref{IJ}) and Eq.~(\ref{Is2}) in the temperature range just below $T_c$.
The network is biased by constant current $I_b$.   The following conditions will be used for the simulation:

\begin{enumerate}
\item The superconductive ground state of each weak link is characterized by current $I_{ij}< \min\{I_{c,n,{ij}}\}=I_{c,min}$ and $V_{ij}=0$. The symmetry is not broken, thus $\tilde{J}_{n,{ij}}$ vanishes.  The conductance of the weak-links in the superconductive state is taken $G \gg e^2/\hbar \,\,\, [\mathrm{\Omega^{-1}}]$, i.e. $G$ is much greater than the quantum conductance $e^2/\hbar$. This condition guarantees the existence of the superconductive ground state.
\item  The intermediate states correspond to the coexistence of superconducting and normal domains. According to the two-fluid model, unpaired electrons coexist with paired electrons in the region of temperature close to $T_c$  respectively, with densities:
      \begin{eqnarray}
        n_\mathrm{N}(T) &=& n_o  \left(\frac{T}{T_c}\right)^4 \\
        n_\mathrm{S}(T) &=& \frac{n_o}{2} \left[1-\left(\frac{T}{T_c}\right)^4\right]
      \end{eqnarray}
           where $n_o$ is the total density of normal electrons. The fraction $n_\mathrm{S}$ of superelectrons is characterized  by critical current $I_{c,n,ij}=\tilde{I}_{n,{ij}}$.  Conversely, the fraction $n_\mathrm{N}$ of normal electrons has a finite value of  $\tilde{J}_{n,ij}$ and, thus, from Eq.~(\ref{IJ}), is characterized by critical current $I_{c,n,ij}=\sqrt{\tilde{I}_{n,{ij}}^2+ \tilde{J}_{n,{ij}}^2}$.
    The condition $\tilde{I}_{n,ij}<I_{ij} < \sqrt{\tilde{I}_{n,{ij}}^2+ \tilde{J}_{n,{ij}}^2} $ holds  in the intermediate state.
 The conductance of the weak-links in the intermediate states varies between $G$ and $G_o=1/R_o$, as a function of temperature, according to the relative fraction of super to normal electrons.
 \item   The normal state is achieved when the voltage $V_{ij}$ across the junction exceeds $V_{c_{ij}}$. The conductance of the weak-links is $G_o=1/R_o$. The current $I_{ij}$ flowing through each weak-link satisfies:
      $I_{ij}>\sqrt{\tilde{I}_{n,{ij}}^2+ \tilde{J}_{n,{ij}}^2}$.
 \end{enumerate}

The superconductor-insulator transition is simulated by solving the system of Kirchhoff equations at varying temperature.
The critical currents  $\tilde{I}_{n,ij}$ and $\tilde{J}_{n,{ij}}$  are assumed to vary on temperature according to the linearized equations
$\tilde{I}_{n,ij}=\tilde{I}_{{o},n,ij} \left(1- {T}/{T_{c}} \right)^\gamma$
and
$\tilde{J}_{n,ij}=\tilde{J}_{{o},n,ij} \left(1- {T}/{T_{c}} \right)^\gamma \hspace{2pt},$
where $\tilde{I}_{o,n,ij}$ and $\tilde{J}_{o,n,ij}$ are the lowest temperature values of $\tilde{I}_{n,ij}$ and $\tilde{J}_{n,{ij}}$ and the exponent $\gamma$ is about $2$ for high-T$_c$ superconductors. Hence, the critical current $I_{c,n,ij}$ depends on temperature according to  $I_{c,n,ij}= I_{c_o,n,ij} \left(1- {T}/{T_{c}} \right)^\gamma$,   with $I_{c_o,n,ij}=\sqrt{\tilde{I}_{o,n,{ij}}^2+ \tilde{J}_{o,n,{ij}}^2}$.  In order to take into account the disorder of the array,   $\tilde{I}_{n,ij}$ and $\tilde{J}_{n,{ij}}$ are taken as random variables, distributed according to  Gaussian
functions with mean values $\tilde{I}_{{o},n}$  and  $\tilde{J}_{{o},n}$ and standard deviations $\Delta \tilde{I}_{{o},n}=\Delta \tilde{J}_{o,n}$.
\par By effect of the temperature
increase and consequent reduction of the critical current, the weak-link with the lowest value of the critical current $I_{c,n,ij}=I_{c,min}$ switches to the intermediate state and, then, becomes resistive when $V_{ij}>V_c$. The resistive transition of the first weak-link has the effect to set the value of the voltage drop across the other weak-links in the same layer. The result is the formation of a layer of  weak-links either in the resistive or in the intermediate state.
As temperature further increases, the critical
current $I_{c,n,ij}$ further decreases. More and more weak-links  gradually
switch from the superconductive to the intermediate state and then to the resistive state.
The term $\tilde{J}_{n,ij}$ acts by increasing the critical current value of the weak-link in the intermediate state in the layers undergoing the transition.
It is worthy to remark that the increase of critical current is relative to the fraction of normal electrons in the mixed states. The onset of $\tilde{J}_{n,ij}\cos(n\phi_{ij})$  is indeed triggered by the elementary resistive transition of the weak-link with the lowest critical current, since it is related to the partly broken pair-symmetry of the weak-links in the intermediate state. It has no effect on the links in the superconductive state, neither on those in the fully resistive state.
\par
\begin{figure}
\centering
\includegraphics[width=7cm, angle=0]{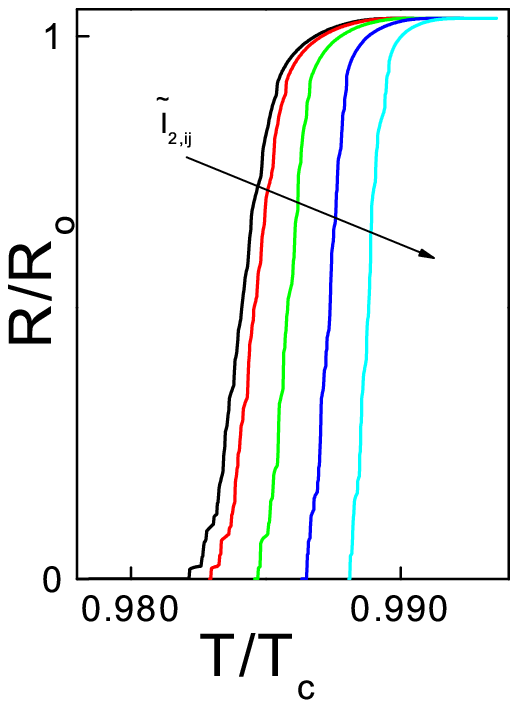}
\caption{\label{Figure3} Resistive transition of a two-dimensional network with current-phase relation of the form $I_{\mathrm{S},ij}(\phi_{ij})=\tilde{I}_{1,ij}\sin(\phi_{ij})+ \tilde{I}_{2,ij}\sin(2\phi_{ij})$. The average value of the critical current $\tilde{I}_{1,ij}$ is $1\mathrm{mA}$.  The curves correspond to different average values of the critical current $\tilde{I}_{2,ij}$, namely $\tilde{I}_{2,ij}=0 \mathrm{mA}$,  $\tilde{I}_{2,ij}=0.5\mathrm{mA}$, $\tilde{I}_{2,ij}=0.75\mathrm{mA}$ and $\tilde{I}_{2,ij}=1\mathrm{mA}$. The normal resistance $R_o$ is $1 \Omega$ equal for all the junctions. }
\end{figure}

Fig.~\ref{Figure3} shows the curves of
the resistive transitions obtained with the current-phase relation $I_{\mathrm{S},ij}(\phi_{ij})=\tilde{I}_{1,ij}\sin(\phi_{ij})+ \tilde{I}_{2,ij}\sin(2\phi_{ij})$ for a two-dimensional $30 \times 30$ network. The curves correspond to different values of the term $\tilde{I}_{2,ij}$. The values of the critical currents are $\tilde{I}_{1,ij}=1\mathrm{mA}$ and $\tilde{I}_{2,ij}$ ranging from  $0$ to $1\mathrm{mA}$. The standard deviation of the  critical currents  is $\Delta I_{c_{o},n}=0.3\mathrm{mA}$.
The effect due to  $\tilde{I}_{2,ij}\sin(2\phi_{ij})$ corresponds to a shift of the transition towards higher or lower temperature depending on amplitude. As opposed to $\tilde{J}_{n,ij}\cos(n\phi_{ij})$, the term $\tilde{I}_{2,ij}\sin(2\phi_{ij})$ acts on the Josephson coupling and thus its effect is higher at the beginning of the transition and decreases as the fraction $n_{\mathrm{S}}$ of superelectrons decreases.
\par
\begin{figure}
\centering
\includegraphics[width=10cm, angle=0]{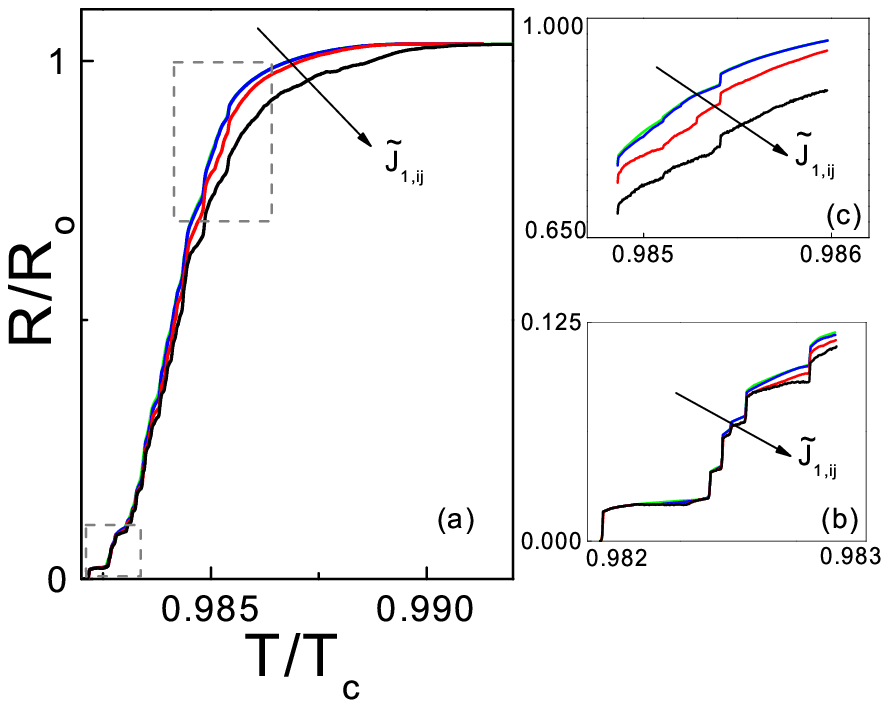}
\caption{\label{Figure4} Resistive transition of a two-dimensional network with current-phase relation of the form $I_{\mathrm{S},ij}(\phi_{ij})=\tilde{I}_{1,ij}\sin(\phi_{ij})+ \tilde{J}_{1,ij}\cos(\phi_{ij})$. The average value of the critical current $\tilde{I}_{1,ij}$ is $1\mathrm{mA}$.
The curves correspond to different average values of the critical current $\tilde{J}_{1,ij}$, namely $\tilde{J}_{1,ij}=0 \mathrm{mA}$, $\tilde{J}_{1,ij}=0.5\mathrm{mA}$, $\tilde{J}_{1,ij}=0.75\mathrm{mA}$ and $\tilde{J}_{1,ij}=1\mathrm{mA}$. The normal resistance $R_o$ is $1 \Omega$ equal for all the junctions. Panels (b) and (c) show the details of the beginning and the end of the transition.}
\end{figure}

Fig.~\ref{Figure4} shows the curves of
the resistive transitions obtained with current-phase relation $I_{\mathrm{S},ij}(\phi_{ij})=\tilde{I}_{1,ij}\sin(\phi_{ij})+ \tilde{J}_{1,ij}\cos(\phi_{ij})$ for a two-dimensional $30 \times 30$ network.
The curves correspond to different values of the term $\tilde{J}_{1,ij}$. The values of the critical currents are $\tilde{I}_{1,ij}=1\mathrm{mA}$ and $\tilde{J}_{1,ij}$ ranging from $0$ to $1\mathrm{mA}$. The standard deviation of the  critical currents is $\Delta I_{{o},n}=0.3 \mathrm{mA} $.
Initially, the weak-links  are in the superconductive state, thus the network resistance is negligible. As temperature increases, the weak-link with the lowest critical current switches to the intermediate state and then to the resistive state with the consequent onset of the term $\tilde{J}_{n,ij}\cos(n\phi_{ij})$ and redistribution of the currents. One can notice that the curves overlap at the beginning of the transition, whereas become more separated when $T \rightarrow T_c$, implying that the effect of the term $\tilde{J}_{n,ij}\cos(n\phi_{ij})$ is more relevant as the transition approaches its end. The amplification of the $\tilde{J}_{n,ij}\cos(n\phi_{ij})$ effect, as the resistance increases, means that $\tilde{J}_{n,ij}$ acts as modulation of the resistance.
The modulation effect due to  $\tilde{J}_{n,ij}$  can be noted at the level of each elementary transition step.
Fig.~\ref{Figure4}(b) and (c) show the zoom of the resistance steps corresponding respectively to the beginning and to the end of the transition curves in Fig.~\ref{Figure4}(a).  One can notice that the microscopic deviations from the staircase profile obtained with the simple sinusoidal current-phase relation increase as the global network resistance increases in agreement with the modulating action of the term $\tilde{J}_{n,ij}$.

\subsection{Critical temperature anomaly in cuprates.}
\begin{figure}
\centering
\includegraphics[width=8cm, angle=0]{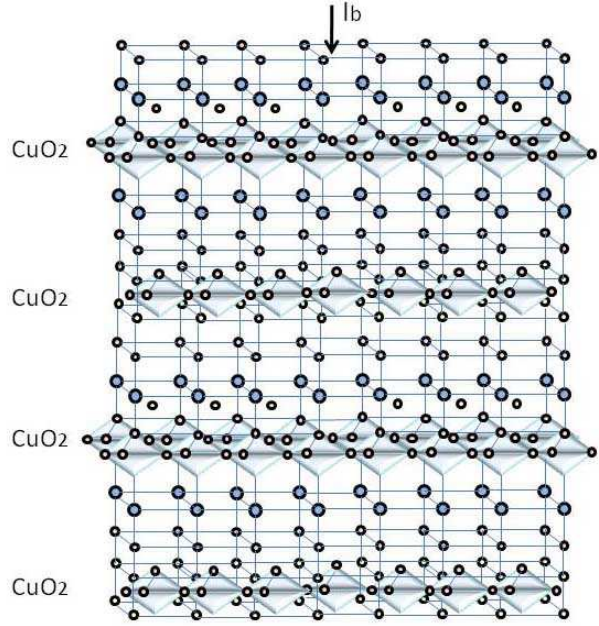}
\caption{\label{Figure5} Arrays of intrinsic Josephson junctions are naturally formed in cuprates by the CuO$_2$ planes separated by layers of insulating atoms. The hole doping $p$ of the CuO$_2$ planes affects transport and thermodynamic properties of cuprates. Several transport anomalies have been observed around $p=1/8$  that cannot be explained in the framework of a conventional picture of the intrinsic Josephson junctions and have been ascribed to the antiphase ordering across the planes [52-58]. The modulation of the phase can be taken into account by using the proposed array of resistively and capacitively nonsinusoidal Josephson junctions.}
\end{figure}

\begin{figure}
\centering
\includegraphics[width=8cm, angle=0]{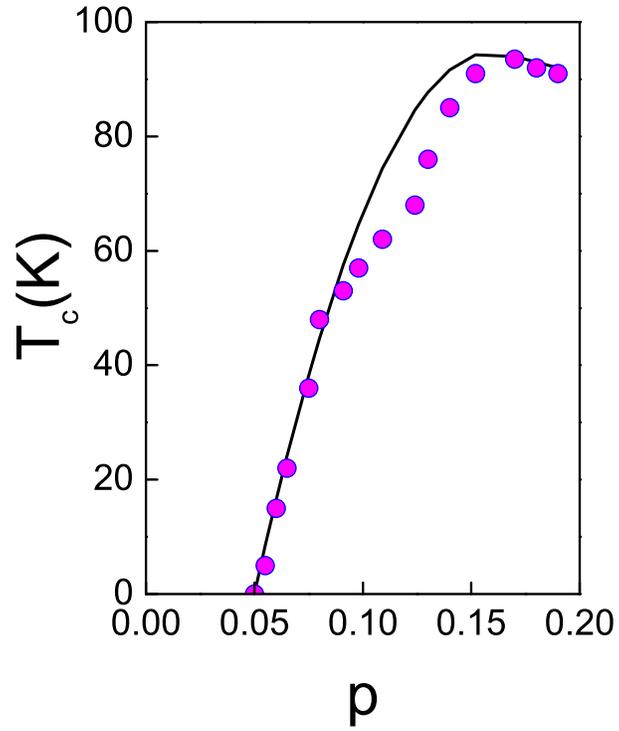}
\caption{\label{Figure6} Critical temperature T$_c$ as a function of the hole doping $p$. The ideal parabolic relation is  plotted as a reference (solid line). Circles are experimental data obtained on YBCO samples with varying doping level of the CuO$_2$ planes (\cite{Liang}). The suppression of T$_c$ in the range of doping between 0.08 and 0.17 can be observed.}
\end{figure}

\begin{figure}
\centering
\includegraphics[width=9cm, angle=0]{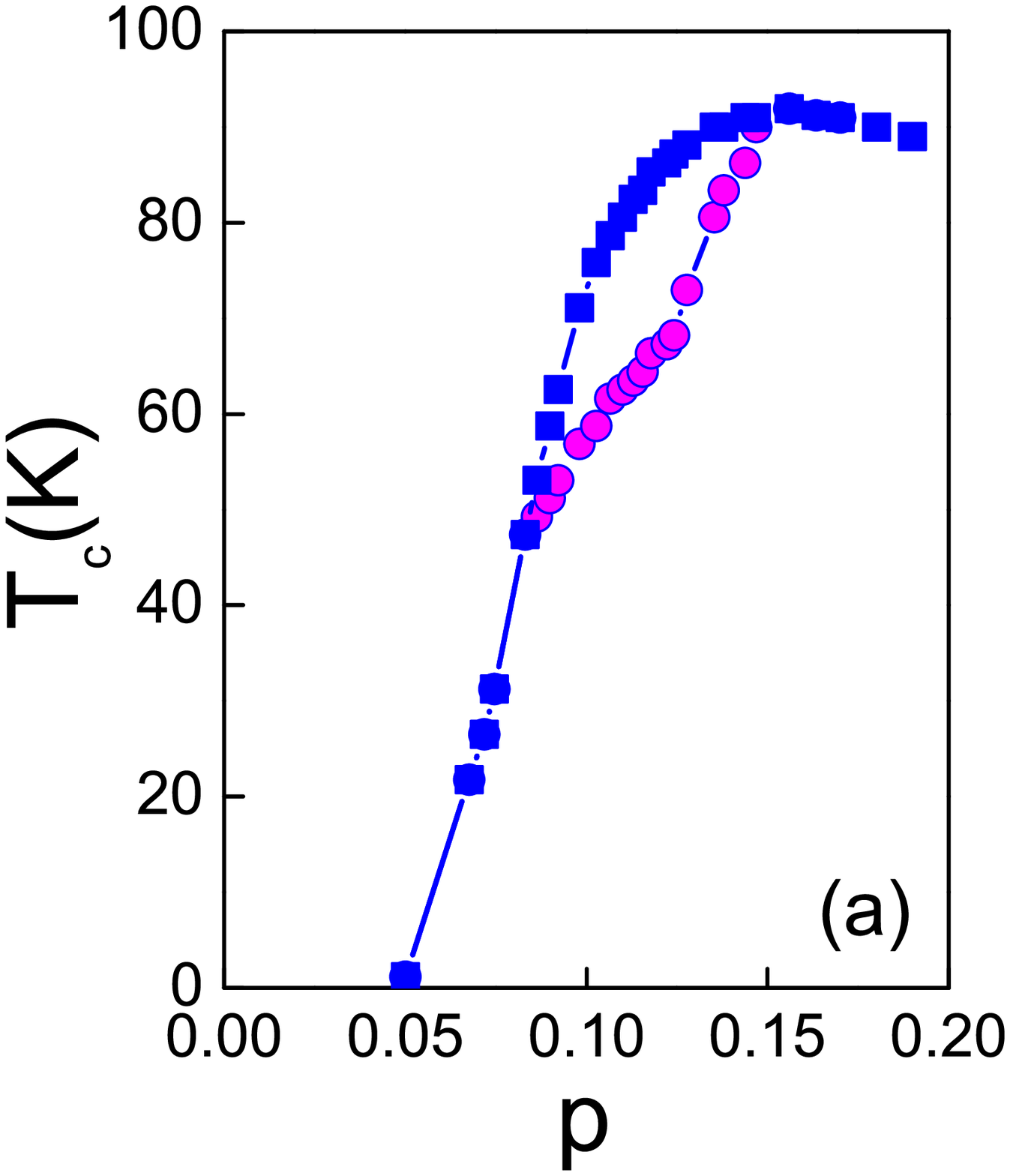}
\includegraphics[width=9cm, angle=0]{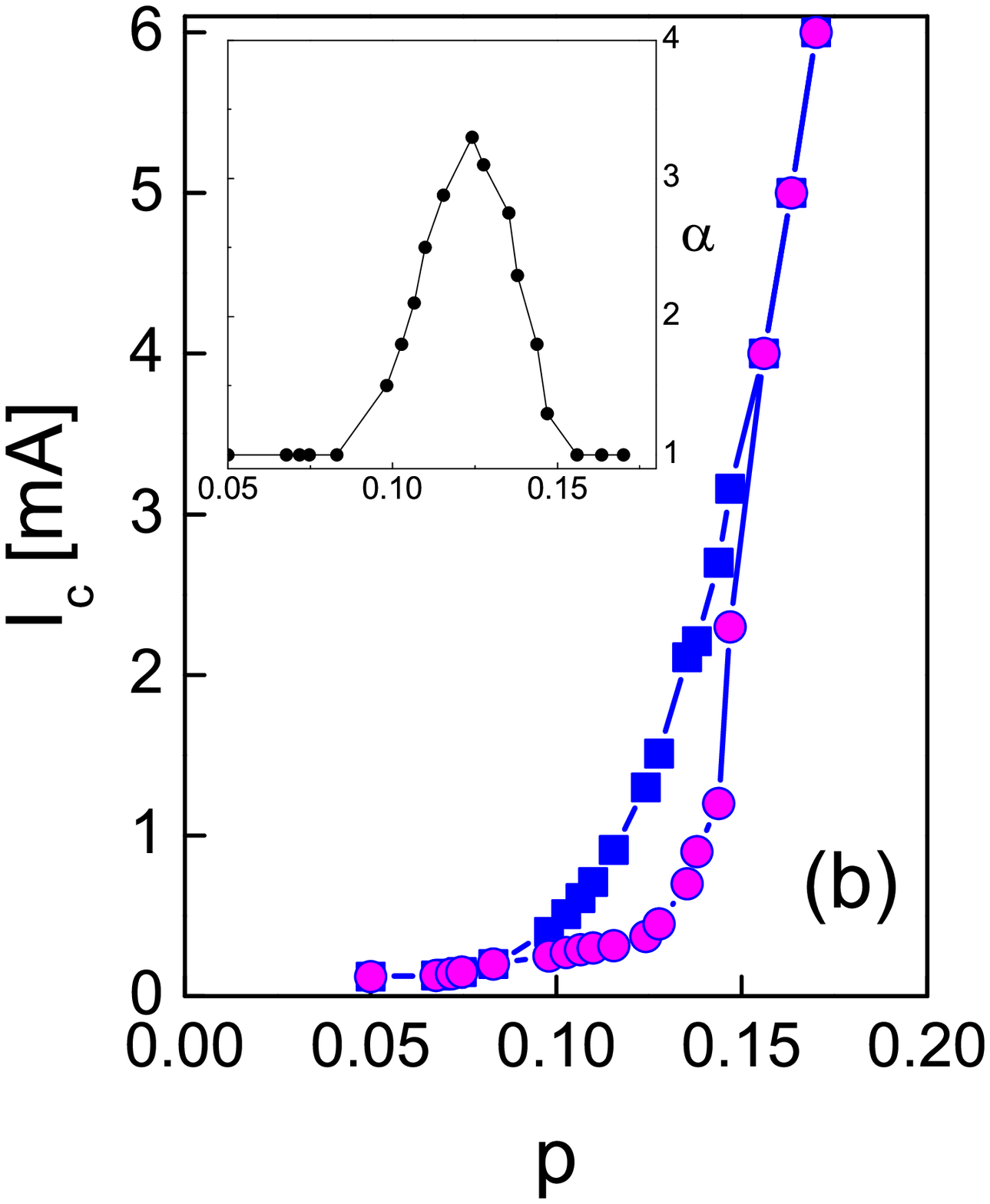}
\caption{\label{Figure7} Critical temperature T$_c$ (a) and currents I$_c$ (b) as a function of the hole doping $p$.   The data are obtained by simulating the network of Josephson junctions  with current-phase relation given by $I_{\mathrm{S},ij}(\phi_{ij})=\tilde{I}_{1,ij}\sin(\phi_{ij})+ \tilde{I}_{2,ij}\sin(2\phi_{ij})$. The average critical current of the array takes values  in the range $1 \, \mathrm{mA}\div \, 10 \mathrm{mA}$. In order to yield the suppression of $T_c$ as a function of the doping level the component $\tilde{I}_{2,ij}$ is reduced. In the inset of (b) the ratio $\alpha$ of the $\tilde{I}_{2,ij}$ components corresponding to  the ideal parabolic behavior and the real curve is plotted.}
\end{figure}

 In the previous section, the approach has been applied to a granular superconductor where the disorder of the material is taken into account by using a suitable probability distribution function of a relevant parameter. In particular, the probability distribution function is a Gaussian with the variance $\sigma$ accounting for the randomness of the critical currents $I_c$ over the array. In this section, the approach is implemented to model perfectly ordered single crystals. In the absence of localized and extended defects, the relevant parameters of the arrays of Josephson junction are expected to be deterministic and thus a probability distribution function is not necessary, i.e. $\sigma = 0$.\\
Specifically, the proposed method will be applied to the intrinsically formed arrays of Josephson junction in single crystal of layered cuprates.
 The reported simulations are mainly addressed to describe the occurrence of a $\pi$-phase shift within the CuO$_2$ planes in accordance with the anti-phase ordering model put forward by Berg et. al. \cite{Berg} to account for many anomalies exhibited by cuprates. The existence of an anti-phase ordering has been experimentally confirmed in \cite{Liang,Yuli,Homma,Naqib}. The approach presented in this work is particularly suitable to simulate the array of intrinsic Josephson junctions with unconventional current voltage characteristics and the onset of $\pi$ phase.
Thus, the goal is a phenomenological description of the model \cite{Berg} and the corresponding simulation of the experimental results presented in \cite{Liang,Yuli,Homma,Naqib}. In particular, the predictions are compared with the experimental data concerning the T$_c$ suppression  observed in cuprates at varying levels of doping. The doping $p$, i.e. the number of holes per copper atom in the CuO$_2$ planes, is a key quantity determining the main
properties of high-T$_c$ superconductors, whose typical structure is shown in Fig.~\ref{Figure5}.
A parabolic relationship between superconducting transition temperature T$_c$  and doping  $p$  has been envisaged:
\begin{equation}\label{Tc}
1-\frac{T_c}{T_{\mathrm{c,max}}}=82.6 \,(p-0.16)^2 \hspace{5pt}.
\end{equation}
This relation is inaccurate for certain values of the doping and a very pronounced T$_c$ suppression (as shown in Fig.~\ref{Figure6}) has been reported in many cuprates \cite{Liang,Yuli,Homma,Naqib}.
Such a universal suppression of  T$_c$  has been ascribed to the tendency of charge stripe formation, with  spatially modulated  superconducting order and phase \cite{Berg}. Evidence of stripe order in cuprates is provided by the enhancement of the anisotropy of resistivity with temperature. The charge dynamics is those of a superconductor in plane at high temperature but the behavior is that of a poor metal in the orthogonal direction.
 At low temperature, the effective Josephson coupling is always positive yielding the homogeneous low-T$_c$ superconducting phase, while the striped superconducting phase is found at relatively higher temperature. The scheme describing such phenomenon is given by alternating stripes of superconductor and insulator, forming an array of Josephson junctions.
For a $d$-wave superconductor at high temperature, with a strong crystal field coupling which locks the lobes of the pair-wave function along the crystallographic axis,  the order parameter may change sign under rotation across the planes. The overall result is an effective higher order Josephson coupling depending  on the cosine of twice the difference of the superconducting phases on neighboring planes and the dominance of a  negative $\sin(2\phi)$ component in the current-phase relation \cite{Berg}.\\
Such an array of Josephson junctions, with spatially modulated Josephson coupling, exhibiting anomalous transport and thermodynamics, could not be accounted for by the resistively and capacitively shunted model with simple sinusoidal current-phase relation. Therefore, in the present work, a simulation based on arrays of nonsinusoidal Josephson junctions is put forward. The scope is the estimation of the critical current components yielding the parabolic dependence  of T$_c$ and the suppression  observed  at doping values close to $p=1/8$. The different doping level of the CuO$_2$ planes is taken into account by varying the critical current, which is related to the number of Cooper pairs in the superconductive phase and thus enhanced/suppressed by the hole doping. Specifically, the variation of the negative component  $\tilde{I}_{2,ij}\sin(2\phi_{ij})$ dominating over the simple sinusoid is taken into account as the origin of the suppression of the critical temperature. As already stated, the simulations  refer to a perfect crystal lattice  instead of a granular superconductor, thus the array is perfectly ordered and  $\sigma$ is negligible ($\sigma = 0$). Hence, one
can expect that the superconductivity is easily suppressed by
current perpendicular to the superconducting layers while current
flowing parallel to the layers would not destroy the superconducting
state of the crystal. In the following, two sets of simulations are performed.\\
First, the transition is simulated  to obtain the ideal parabolic dependence given by Eq.~(\ref{Tc}). The differential equation (\ref{I1}) is solved for nonsinusoidal junctions with $\tilde{I}_{1,ij}$ and  $\tilde{I}_{2,ij}$ components. The I-V characteristics of the single junctions is obtained and then implemented to simulate the transition of the whole array as described in the previous sections.  The transition  curve  allows one to deduce the critical temperature  by  using the relation $\tilde{I}_{n,ij}=\tilde{I}_{{o},n,ij} \left(1- {T}/{T_{c}} \right)^\gamma$, where the exponent  $\gamma$ is taken equal to 2. In the present work, the critical currents  have been varied in the range $0.1\,\div \,10 \mathrm{mA}$ corresponding to doping level $p$ varying between $0.05 \,\div \, 0.18$ and critical temperature varying between $0\,\div \, 95 \mathrm{K}$ according to the data of Refs.\cite{Liang,Yuli,Homma,Naqib}.  The critical temperatures and currents obtained from the simulation are shown in Fig.~\ref{Figure7} (a) and (b) (blue squares).\\
Then, the transition is simulated to obtain the suppression of T$_c$ with respect to the parabolic dependence. The suppression of T$_c$ is obtained by a decrease of I$_{2,ij}$  as a function of the doping level for values ranging from $p=0.08$ to $p=0.17$. The critical temperatures and currents obtained from the simulations are shown in Fig.~\ref{Figure7} (a) and (b) (magenta circles).\\
In the inset of Fig.~\ref{Figure7}(b) the ratio $\alpha$ of the second harmonics  for the ideal parabola and the real curve with suppression, is plotted. The maximum temperature suppression corresponds to a value of the ratio close to 3.5.

\section{Conclusions}
The nonsinusoidal  current-phase relation has been considered in the resistively shunted Josephson junction model  for describing the superconductive transition. By solving a system of Kirchhoff equations for the array of nonsinusoidal Josephson junctions, it is found that additional cosine and sine terms modify the transition curves by changing  resistance  and Josephson coupling in the framework of the two-fluid model of superconductivity.
 Higher harmonics, besides the simple sinusoid $I_{c}\sin\phi$, might arise in the vicinity of the transition because of the  nonstationary conditions and the random distribution of potential drops and impurities in granular superconductors.
The approach has been implemented for characterizing the critical temperature suppression observed in cuprates.  In particular, our focus is on the  anomalies experimentally observed in cuprates that need to go beyond the simple sinusoidal picture arising from a constant positive Josephson coupling valid for low-T$_c$ superconductors.
The specific example of the $T_c$ suppression at doping level $p=1/8$ is described in terms of the ratio of  the second to first sinusoidal components of the current-phase relation. The  naturally formed networks of Josephson junctions, due to insulating layers sandwiched between CuO$_2$ planes, are strongly affected by the presence of higher-order terms in the current phase relation.  Further applications of the present approach can be envisaged to account for  the complex  phenomenology of high-T$_c$ materials forming arrays of nonsinusoidal Josephson junctions and its implications in novel quantum devices.

\par

\section{acknowledgements}
The \emph{Istituto Superiore per le Telecomunicazioni M. Boella}  is gratefully acknowledged for financial support.


%

\clearpage
\clearpage

\clearpage
\clearpage
\clearpage
\clearpage
\clearpage

\end{document}